\documentclass{tMPH2e}
\usepackage{graphicx}
\usepackage{amsmath}

\newcommand*{\vect}[1]{\ensuremath{\bm #1}}
\newcommand*{\matr}[1]{\ensuremath{\bf #1}}

\begin{document}

\title{Simulating rotationally inelastic collisions using a Direct Simulation Monte Carlo method}
\author{O. Schullian$^{a,d}$, J. Loreau$^b$, N. Vaeck$^b$, A. van der Avoird$^c$, B. R. Heazlewood$^d$, C. J. Rennick$^d$, and T. P. Softley$^{d\ast}$\thanks{$^\ast$Corresponding author. Email: tim.softley@chem.ox.ac.uk}\\%
    \vspace{6pt}%
    $^a$\em{Laboratorium für Physikalische Chemie, ETH Zürich, CH-8093 Zürich, Switzerland}; %
    $^b$\em{Service de Chimie Quantique et Photophysique, Universit\'e Libre de Bruxelles (ULB) CP 160/09, 50 av. F.D. Roosevelt, 1050 Brussels, Belgium}; %
    $^c$\em{Theoretical Chemistry, Institute for Molecules and Materials, Radboud University, Heyendaalseweg 135, 6525 AJ Nijmegen, The Netherlands}
    $^d$\em{Department of Chemistry, University of Oxford, Chemistry Research Laboratory, Mansfield Rd, Oxford, United Kingdom}
}

\maketitle

\begin{abstract}
A new approach to simulating rotational cooling using a direct simulation Monte Carlo (DSMC) method is described and applied to the rotational cooling of ammonia seeded into a helium supersonic jet. 
The method makes use of {\it ab initio} rotational state changing cross sections calculated as a function of collision energy. Each particle in the DSMC simulations is labelled with a vector of rotational populations that evolves with time.  Transfer of energy into translation is calculated from the mean energy transfer for this population at the specified collision energy.  The simulations are compared with a continuum model for the on-axis density, temperature and velocity; rotational temperature as a function of distance from the nozzle is in accord with expectations from experimental measurements. The method could be applied to other types of gas mixture dynamics under non-uniform conditions, such as buffer gas cooling of NH$_3$ by He.
\end{abstract}


\section{Introduction}

Rotational cooling of molecules in supersonic jets or in cryogenic buffer gas cells is a key component of many spectroscopic and collision  dynamics experiments.
In general, molecules seeded into an appropriate rare-gas jet undergo efficient inelastic collisions in the isentropic supersonic expansion leading to rotational temperatures in the beam on the order of a few kelvin. The resulting simplification of molecular spectra originating from just a few populated quantum states has been extremely valuable to the gas-phase spectroscopic community.  It has also allowed the production of quantum-state selected beams for controlled collision experiments.
Translationally, the supersonic molecular beam is cold in the moving frame of reference, however the beam is typically moving at a speed around 500--1000~m~s$^{-1}$.

More recently, there has been experimental interest in extending molecular control to the translational degrees of freedom in the lab frame. The use of rotationally cold supersonic beams as a source for Stark and Zeeman decelerators has led to a number of novel collisional experiments \cite{meijer,merkt,osterwalder,Heazlewood2015-Low-Temperature-Kinetics-and-Dynamics} with both translational and rotational degrees of freedom cooled. The interaction potentials of reacting polyatomic molecules are anisotropic, so the rate will be dependent on the initial rotational level \cite{Dubernet1989-Rotational-state-dependence}.
The molecular anisotropy influences the potential, so excess rotation can prevent the reaction complex locking-in to a favourable orientation. As a consequence, for example, 
the rate of ion-polar neutral reactions decreases with increasing rotational quantum number \cite{Clary1987-Rate-constants-for-the-reactions}.

Buffer-gas cooling methods exploit the inelastic collisions between the sample species and a cold bath gas to remove internal and translational energy \cite{Hutzler2012-The-Buffer-Gas-Beam:-An-Intense}.
In such a setup, the hot molecular species of interest is mixed with an inert buffer gas in a small cryogenic cell, cooled to $\sim 4$~K.
The density is kept sufficiently low that cluster-forming three-body collisions are minimised, and the molecular species undergoes collisions only with the buffer gas before exiting the cell, avoiding collisions with the walls of the cell that are below the freezing point of the molecule.
A beam of slow, cold molecules leaves the cell through a small aperture, and it can be used as a free jet, or it can  be steered and velocity-filtered using a quadrupole or hexapole guide\cite{Sommer2009-Continuous-guided-beams,Twyman2014-Production-of-cold-beams}. 

In order to optimise the conditions for producing rotationally and translationally cold beams, it would be valuable to be able to simulate the rotational and translational cooling processes---for example, the properties of a beam emerging from a buffer gas cell are critically dependent on the shape and dimensions of the buffer gas cell.
The flow of matter and energy in non-equilibrium rarefied gases can be described by the Boltzmann equation, but this is intractable for application to all but the most basic systems.
Instead, the evolution of the gas flow is determined by numerical integration of equations describing the motion of the particles and their collisions.
Direct simulation Monte Carlo (DSMC), also known as Bird's method, is an approach commonly used to model elastic collisions, in which sample particles are taken to represent the distribution functions of a much larger number of real gas particles \cite{Bird1994-Molecular-Gas-Dynamics-and-the-Direct}.
The gas flow is simulated by separating the particle motion and collisions; the position is updated deterministically by the particles' velocity, then collisions are handled stochastically, and post-collision velocities determined.

The Larsen-Borgnakke model is often used to model inelastic collisions in DSMC \cite{Borgnakke1975-Statistical-collision-model}.
In that approach, the major fraction of the collisions is considered to be elastic, and only the post-collision velocities are recomputed.
The small fraction of inelastic collisions is computed statistically, where the post-collision internal and translational energies are sampled from a thermal distribution.
This phenomenological procedure requires selection of an appropriate inelastic collision probability and relaxation rate to match experimental observations.
Such models assume a statistical energy distribution, defining only a rotational temperature, an assumption that breaks down under strongly non-equilibrium conditions \cite{Wysong1998-Assessment-of-direct-simulation}.
Experimentally,  the cooling process may not reach equilibrium if it occurs over a limited time scale and a finite number of collisions, or population may become trapped in metastable states due to symmetry considerations of allowed rotational transitions. 
Improved simulations include rotational energy as a continuum \cite{Tokumasu1999-Dynamic-molecular-collision} or as discrete levels \cite{Boyd1993-Relaxation-of-discrete-rotational,Koura1992-Statistical-inelastic-cross-section}, where the cross section for collision-induced rotational transitions is determined from a rigid rotor model.

In this paper, we report the implementation of an  extension to DSMC that is capable of simulating changes of rotational populations due to inelastic collisions in gas-mixtures of non-uniform density. The method makes use of the explicit collision-energy-dependent state-to-state cross sections for ammonia-helium collisions, calculated using a four-dimensional potential energy surface (PES)---three He-atom spherical coordinates, and the NH$_3$ umbrella angle. The method is applied here to simulate rotational cooling in a supersonic beam, but equally could be employed to simulate rotational cooling in a buffer gas cell.

\section{DSMC model}

The (DSMC) method, developed by Bird in the 1960s, is a well-established tool for simulation of gas flows over a wide range of conditions \cite{Bird1994-Molecular-Gas-Dynamics-and-the-Direct}.
This model numerically integrates the Boltzmann equation for the position and velocity of a collection of simulated particles, each of which represents a large number of physical particles.
The position of each simulated particle is updated by free-flight---independently  of the position of other particles---over a short time interval $\Delta t$, which is less than the mean time between collisions.
The simulation volume is divided into subcells, with dimensions of the order of the mean free path, 
containing (at least for some subcells)
pairs of candidate particles considered for collision. 
The probability of a collision between a pair of simulated particles during the interval $\Delta t$ is equal to the ratio of the volume swept out by their cross sections to the volume of the subcell. In the case that the collision is accepted between a He atom and NH$_3$ molecule, the collision energy is calculated, and the appropriate rotationally inelastic cross section is used as described below. The randomly-oriented post-collision velocity is then determined, taking energy transfer into account, and the next timestep of the calculation is performed.
 
In this work, collisions between pairs of He atoms are treated at the hard-sphere level of approximation, and NH$_3$-NH$_3$ collisions are assumed to be purely elastic such that there is no rotational state change on collision; this is justified given  that the most probable outcome of these inelastic collisions is resonant energy transfer as the molecules exchange rotational state.

\subsection{Scattering cross sections for ${\text NH}_3 + {\text He}$}

The NH$_3$--He system has been the subject of numerous previous studies in which the interaction potential \cite{Hodges2001-Intermolecular-potential-for-the-interaction,Gubbels2012-Scattering-resonances-in-slow} and the cross sections for rotationally inelastic collisions \cite{Green1976b,Machin2005-Rotational-excitation-and-de-excitation,Yang2008-Rotational-excitation-of-NH3-and-ND3-due-to-He-atom,Gubbels2012-Scattering-resonances-in-slow} were computed.
In recent work, state-to-state cross sections were measured at a collision energy of 430 cm$^{-1}$ in hexapole-selected molecular beams \cite{Tkac2014-State-to-state-resolved-differential}, and a theoretical investigation of scattering in the low energy regime ($E<100$ cm$^{-1}$), corresponding to typical kinetic energies available in a Stark-decelerated beam, was carried out for the initial state $1_1^-$ \cite{Gubbels2012-Scattering-resonances-in-slow}.

The cross sections for translational-to-rotational energy transfer calculated in this work are based on the potential energy surface (PES) of the NH$_3$--He complex developed by Gubbels \emph{et al.}, which was fitted from 4180 {\it ab initio} points calculated at the CCSD(T) level of theory \cite{Gubbels2012-Scattering-resonances-in-slow}. It is a four-dimensional PES, which depends on the three spherical coordinates of the He atom relative to the NH$_3$ centre of mass ($R$, $\theta$, $\phi$), and the NH$_3$ inversion motion, described by the umbrella angle $\rho$. The N-H bond length is frozen at the vibrational ground state average value.

The rotational levels of NH$_3$ are labelled by the quantum numbers $J$, $K$, $\pm$. $J$ is the rotational angular momentum quantum number, while $K$ is its projection on the symmetry axis. Each rotational level is split by inversion tunneling into a doublet separated by about 0.79 cm$^{-1}$. The two doublet components are represented by $+$ or $-$, corresponding respectively to symmetric and antisymmetric tunneling states.
The NH$_3$ molecule can exist in two nuclear spin configurations: levels where $K$ is a multiple of 3 or zero correspond to $A$ symmetry and are referred to as {\it ortho}-NH$_3$, and levels where $K$ is not a multiple of 3 are of $E$ symmetry and are referred to as {\it para}-NH$_3$.
Levels corresponding to different symmetries do not interconvert during a collision, so the cross section for such collisions is zero.
Other transitions are allowed with a particular cross section, and must follow energy conservation.

The state-to-state cross sections were calculated by means of the close-coupling method for inelastic scattering of a symmetric top with an atom, which is described in detail in the literature \cite{Green1976b, Gubbels2012-Scattering-resonances-in-slow}. The ammonia inversion was treated explicitly by including the kinetic and potential energy operators corresponding to the umbrella motion in the Hamiltonian of the system \cite{Gubbels2012-Scattering-resonances-in-slow}. In our calculations, we kept the four lowest vibration-inversion states of the umbrella motion. The cross sections were then obtained by solving the close-coupling second-order differential equations with the renormalised Numerov propagator. 
We calculated the state-to-state cross sections for collisional energy transfer between all symmetry- and energy-permitted rotational levels up to $J_K^\pm = 8_8^-$ on a fine grid of collision energies between 0.01~cm$^{-1}$ and 150~cm$^{-1}$, corresponding to the experimental velocities. Owing to the large number of levels under consideration, and the expected population of the higher-lying levels at the low temperature  in a supersonic expansion, the maximum initial level considered here is $8_8^-$.
For each energy, the cross sections were converged with respect to the size of the rotational basis set, the total angular momentum quantum number, the size of the radial grid and the number of grid points.

\begin{figure}[htb]
    \includegraphics{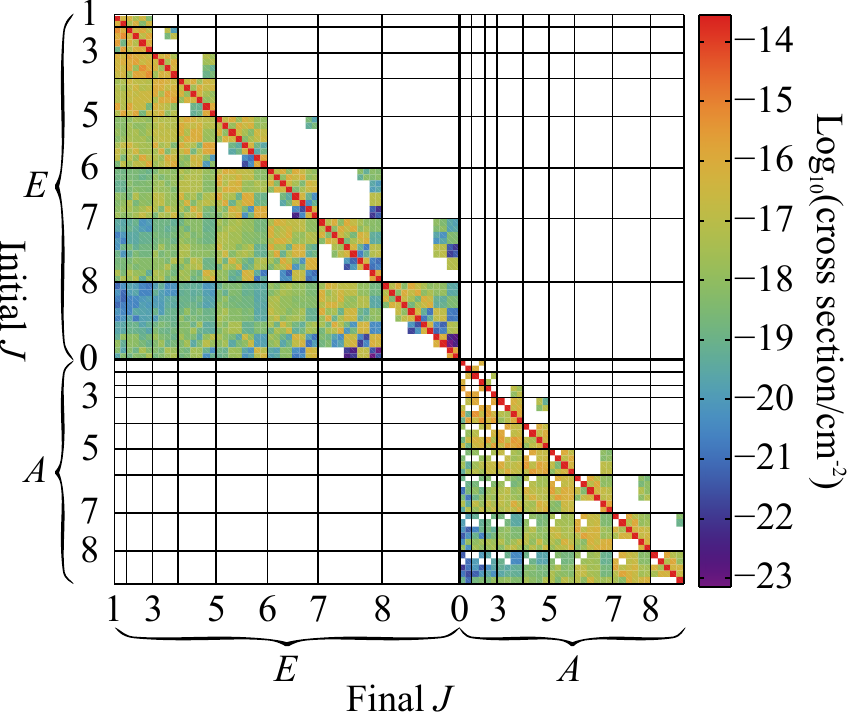}
    \caption{Cross section for collision-induced rotational energy transfer at a collision energy of 25 cm$^{-1}$ from an initial state on the $x$ axis and a final state on the $y$ axis, up $J_K^\pm=8_8^-$. The plot is split into quadrants of $A$ and $E$ symmetry, corresponding to $K=0$ or multiple of 3 ({\it ortho}-NH$_3$), and $K$ not a multiple of 3 ({\it para}-NH$_3$), respectively. White represents  cross section values of zero.}
    \label{fig:crossSections}
\end{figure}

An example of the state-to-state cross sections is shown in Figure \ref{fig:crossSections} in matrix form for a collision energy of 25 cm$^{-1}$. Elastic scattering strongly dominates, as can be expected from the low anisotropy of the PES.

\subsection{Rotationally inelastic collisions}

Each simulated NH$_3$ particle represents a large number of physical particles, so it is assigned a normalised vector of rotational level populations $\vect{\Pi}$.
The number of rotational levels is unbounded, requiring an infinite probability matrix in principle; however in this work we truncate the matrix to a maximum rotational level.
At the temperatures considered here, most of the population lies in the lowest rotational levels, and the low collision energy results in a higher probability for rotational de-excitation than excitation; the population will tend to redistribute to lower energy states.
The evolution of this vector with each successive collision is described by a Markov chain, which represents a stochastic process defined at integer values of time, and the state at each step depends only on the state in the previous step.

At each time step, the probability for the transition from an initial state to a final state is represented by a probability matrix $\matr{P}$, a square matrix of positive quantities in which the sum of the elements of each row is 1.
The relative post-collision rotational population distribution $\vect{\Pi}^f$ is given by the matrix product of the initial distribution $\vect{\Pi}^i$ and the probability matrix $\matr{P}$,
\begin{equation}
    \Pi_j^f = \sum_k{\Pi_k^i P_{kj}} .
\end{equation}
\textbf{The energy-dependent probability of a collision-induced transition from state $i$ to state $j$ is defined as the inelastic cross section $\sigma_{ij}$ divided by the total cross section $\sigma_{\text{tot}}$. The total cross section for the initial state $i$ is the sum of all cross sections for that state, including elastic collisions, $\sigma_{ii}$}
\begin{equation}
    P_{ij}(E) = \frac{\sigma_{ij}(E)}{\sigma_{\text{tot}}(E)} = 
    \frac{\sigma_{ij}(E)}{\sum_k \sigma_{ik}(E)}
    \label{equ:probmatrix}
\end{equation}

Repeated application of the probability matrix generated from collision cross sections to a populations vector converges, yielding an estimate of the rotational cooling rate when combined with the collision rate.
The distribution after two collisions is represented by $\matr{P}\cdot\matr{P} = \matr{P}^2$, and logically must converge to equilibrium for infinite number of collisions \textbf{to satisfy detailed balance}: $\lim_{n\rightarrow\infty}\matr{P}^n = \matr{P}^{\text{eq}}$.
\textbf{The inelastic cross sections satisfy detailed balance conditions as they are computed from the symmetric $S$ and $T$-matricies ($T=I-S$).}
This corresponds to a Boltzmann population distribution according to the temperature, as under these conditions $\vect{\Pi}^{\text{eq}} = \vect{\Pi}^{\text{eq}}\matr{P}^{\infty}$.
Note that the transition matrix does not represent a completely ergodic process; not all states can interconvert owing to the restriction of transitions to within the same symmetry group.
The sub-matrix representing each symmetry is, however, ergodic, so the final population distribution within a symmetry does not depend on the initial conditions, but the relative total population of each does not change.

The probability of a collision between the simulated particles representing He and NH$_3$ is determined to a first approximation using \textbf{a} hard-sphere particle radius \textbf{of 100~pm; if the collision is chosen, the total cross section is calculated to determine the collision probability for the collision pair.}
The relative collision energy is used to interpolate the cross section calculated from the He-NH$_3$ PES, and construct the probability matrix \eqref{equ:probmatrix}.
Multiplication of this matrix by the rotational population vector yields the post-collision state distribution $\vect{\Pi}^{f} = \vect{\Pi}^{i}\matr{P}$.
Total energy must be conserved, so the post-collision velocities are computed accounting for the rotational to translational energy transfer.
The mean energy released during the collision for the pre-collision population $\vect{\Pi}^i$  is given by $\Delta E_{\text{rot}} = (\vect{\Pi}^{f} - \vect{\Pi}^{i})\cdot \vect{\epsilon}$, where $\vect{\epsilon}$ is a column vector of the rotational energies, relative to the lowest level.
This mean energy difference is used to correct the mean post-collision relative velocity, and the final velocity vectors of the collision partners are rotated through random angles in spherical polar coordinates. This use of the mean energy difference in this way provides an efficient means to simulate the translational energy changes due to inelastic collisions.

\section{Analytical molecular beam model for comparison with simulations }

For seeded supersonic beams in which the density of the seed gas is much less than that of the carrier gas, the collision rate between carrier gas atoms is much greater than with the seed species.
Under these conditions, the seed has a negligible contribution to the expansion, and the mixture can be considered  as a nearly-pure flow of the carrier gas.
The properties of the beam depend on the reservoir temperature and pressure, the mole fraction-weighted particle mass $\bar{m}$, and heat capacity ratio $\bar{\gamma} = \bar{c_p}/\bar{c_v}$.

The gas mixture starts 
in a reservoir at an initial pressure $p_0$ and room temperature $T_0$.
The pressure drop towards the aperture causes the gas to accelerate, reaching sonic speed at the throat of the nozzle for a sufficiently large pressure difference.
The beam continues to accelerate as collisions convert the random thermal motion to directed motion, and the jet asymptotically approaches a terminal velocity $v_\infty$.
As the gas flow speed is greater than the local speed of sound, information in the form of compression waves can not propagate upstream, so the gas flow is independent of the downstream conditions.
The flow is supersonic up to the location of a shock wave, known as the Mach disc, where the flow adapts to the background pressure $p_b$ and becomes subsonic.
This structure is located on the centre-line of the flow at a distance $0.67 d \sqrt{p_0/p_b}$ from the exit of a nozzle of diameter $d$; of the order tens of meters under typical experimental conditions \cite{scholes}.

The continuum-flow thermodynamic analysis of the expansion gives the variation of velocity and density with the Mach number as\cite{scholes}
\begin{align}
    v &= M\sqrt{\frac{\bar{\gamma}R T_0}{\bar{m}}}
    \left(1 + \frac{\bar{\gamma}-1}{2}M^2\right)^{-\frac{1}{2}}
    \label{equ:velocity}\\
    \frac{n}{n_0} &= \left(1 + \frac{\bar{\gamma}-1}{2}M^2\right) 
    ^{-\frac{1}{\bar{\gamma}-1}}
    \label{equ:density}
\end{align}
These relations describe the beam conditions in terms of the source density $n_0$, and a distance-dependent Mach number.
The Mach number is defined as the ratio of the gas speed to the local speed of sound $M=v/a$, and has an analytical form solved by the \emph{method of characteristics} from the partial differential equations that describe the conservation of mass, momentum and energy along a flow streamline. 
Within the supersonic part of the flow, all streamlines appear to radiate from a spherical source located a distance $z_0$ from the nozzle with aperture diameter $d$, and the Mach number as a function of distance from the nozzle $z$ is well-approximated by
\begin{equation}
M = a \left(\frac{z-z_0}{d}\right)^{\bar{\gamma}-1} 
- \frac{1}{2} \frac{\frac{\bar{\gamma}+1}{\bar{\gamma}-1}}
{a \left(\frac{z-z_0}{d}\right)^{\bar{\gamma}-1}}.
    \label{equ:mach}
\end{equation}
The parameters $a$ and $z_0$ are obtained by fitting to experimental data, and depend on the heat capacity ratio $\gamma$; Ashkenas and Sherman tabulate these values for a range of $\gamma$, and the values used here are interpolated for the mean heat capacity ratio of the seeded expansion
\cite{Sanna2005}.

\section{DSMC model results for a supersonic expansion of NH$_3$ seeded in He}

The supersonic expansion modeled here is  a useful  test of the algorithm under the high-density regime of a non-equilibrium flow.
We simulate the expansion of a gas jet into vacuum from a 5~bar, 298~K reservoir through a 0.5~mm orifice.
The model consists of a $5\,\text{mm}\times5\,\text{mm}\times3\,\text{mm}$ volume, with the jet propagating along the $z$ axis from a 0.5~mm disc source positioned centrally on the $5\,\text{mm}\times5\,\text{mm}$ wall; all other walls are treated as vacuum such that particles hitting these are removed from the simulation.
The simulation volume is divided into $400\times400\times240$ cubic sub-cells of dimension 12.5~$\mu$m.
Each simulated particle represents $10^{10}$ physical He or NH$_3$ gas particles, and is introduced into the simulation from the disc source with initial velocities---and rotational population distribution in the case of NH$_3$---chosen from an appropriate thermal distribution.
The simulation propagates the particle positions and velocities in 1~ns time steps, adding particles at a rate appropriate to the reservoir temperature and pressure, and the 1\% seed ratio.
The simulation reaches steady-state when the particle number remains approximately constant with time: the number of particles leaving the volume is equal to the number added.
At this point, data from each subcell are recorded for later analysis: the  total number of particles and the mean velocity for each species, and the mean rotational population distribution.

\subsection{Velocity variation along the beam axis}

\begin{figure}[htb]
    \includegraphics{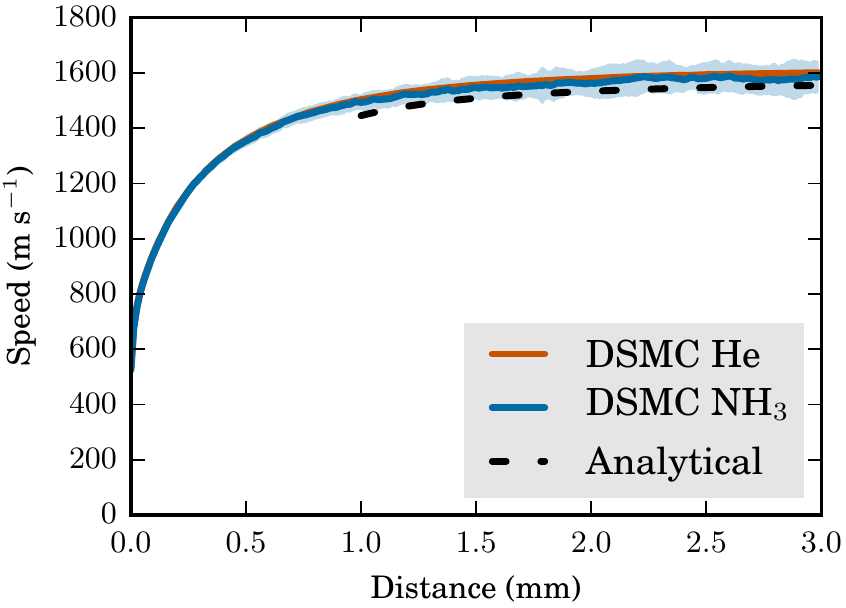}
    \caption{The solid lines and shaded region give the mean velocity and its standard deviation as a function of distance from nozzle exit over a central $125\,\mu\text{m}\times125\,\mu\text{m}$ region of the DSMC model for each species. The lines overlap and are nearly indistinguishable. The dotted line is the analytical expression of velocity for the expansion conditions, and has a region of validity at $z>1.5\,\text{mm}$.}
    \label{fig:velocity}
\end{figure}

Figure \ref{fig:velocity} plots the average and standard deviation of the $z$ component of the velocity for each particle over the central $10\times10$ square cross-section grid of sub-cells, corresponding to a $125\,\mu\text{m}\times125\,\mu\text{m}$ cross section on the centre-line of the jet.
The jet speed predicted by Equation \eqref{equ:velocity} is overlaid as a dotted line, with mean mass and heat capacity ratio of the component species, weighted by the 1\% seed fraction used in the DSMC model.

The DSMC simulations clearly reproduce the jet acceleration near the nozzle exit as random thermal motion is converted to directed motion.
The terminal velocity of each species (NH$_3$ and He) is approximately equal, as collisions are sufficient to entrain and accelerate the ammonia in the expansion, which is dominated by the helium properties.

The DSMC model shows excellent agreement with the analytical model over the range of validity of that model, which is limited by the expression for the Mach number \eqref{equ:mach} that holds for $M>5.5$, which is reached at $z=1.5\,\text{mm}$ under the conditions considered here.
\textbf{The analytical line shown in figure \ref{fig:velocity} lies within the shaded area of the plot, representing one standard deviation of the simulated particles' velocity.}
The remaining overestimate of the terminal velocity can be attributed to the small amount of rotational to translational energy exchange during collisions, which is within the standard deviation of the velocity fluctuations of the DSMC model.

\subsection{Density variation}

\begin{figure}[htb]
    \includegraphics{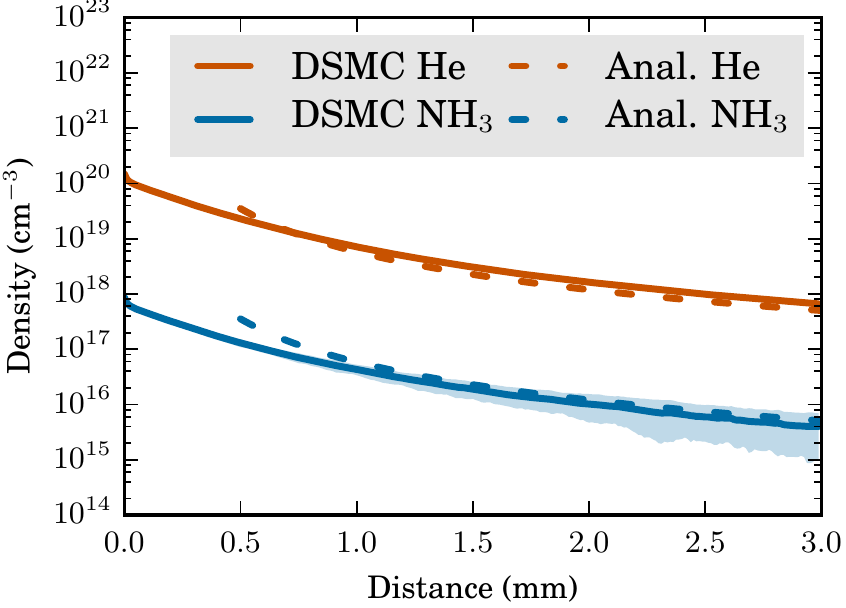}
    \caption{Density of the seed and carrier gases along the centre of a supersonic expansion, predicted by the DSMC model. Upper red line: helium carrier gas, lower blue line: seed ammonia. The shading represents the standard deviation of the density fluctuation over a $125\,\mu\text{m}\times125\,\mu\text{m}$ region on the centre-line of the jet. The dotted lines are the density predicted by equation \eqref{equ:density}, scaled by the reservoir mole fraction of each gas.}
    \label{fig:density}
\end{figure}

The number of particles representing each species in a subcell is converted to a predicted gas density by multiplying each value by the number of physical particles represented by a simulated particle, and dividing by the volume of a subcell.
Figure \ref{fig:density} plots the average of this predicted number density, and its standard deviation, over the same $10\times10$ subcell area along the jet centre line as the velocity.

\begin{figure}[htb]
    \includegraphics{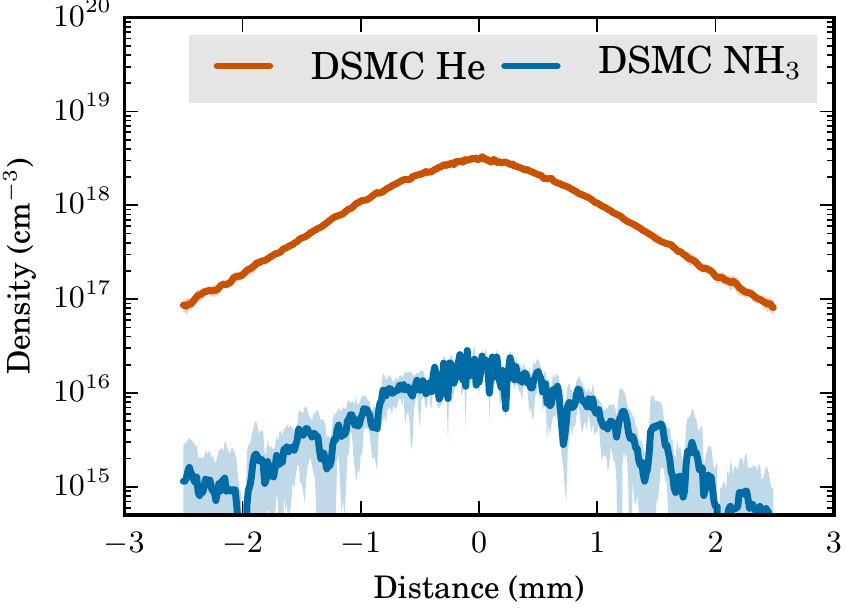}
    \caption{\textbf{Mean seed and carrier gas densities predicted by the DSMC model along a line perpendicular to the propagation direction located 1.5~mm from the nozzle. The shading represents the standard deviation over a $125\,\mu\text{m}$ region.}}
    \label{fig:transverseDensity}
\end{figure}

Equation \eqref{equ:density} predicts the total gas density of a single-species expansion from a given reservoir pressure, so this density is multiplied by the same 1\% NH$_3$ seeding ratio of the source gases for comparison to the density predicted by DSMC.
This scaled density is plotted as the dashed line of matching colour.
As with the velocity variation, there is quantitative agreement between the DSMC model and that calculated analytically, within the range of the analytical hydrodynamic model. The three-dimensional model also correctly captures the Gaussian-shaped off-axis density distribution, shown in figure \ref{fig:transverseDensity} at 1.5~mm from the nozzle.

\subsection{Temperature variation}

\begin{figure}[htb]
    \includegraphics{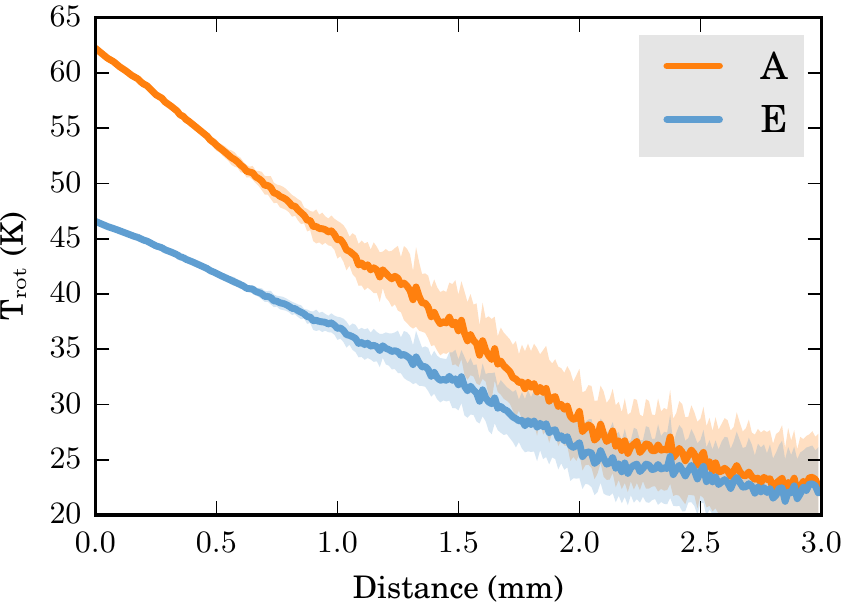}
    \caption{Ammonia rotational temperature estimated from a Boltzmann distribution as a function of distance from the nozzle exit for the rotational levels belonging to each symmetry manifold. \textbf{The shading represents the standard deviation over a central $75\,\mu\text{m}\times75\,\mu\text{m}$ region of the DSMC model.}}
    \label{fig:rotationalTemperature}
\end{figure}

The DSMC simulation explicitly follows the rotational population distribution up to $J_K = 8_8$ for each simulated particle,  a total of 90 levels including the inversion doublet.
The mean population for each level is taken for all particles in a subcell, and recorded after the DSMC simulation has reached steady state.
The rotational temperature is defined by comparing the relative population of each level with a Boltzmann distribution.
Plotting the natural logarithm of the population of each level divided by its degeneracy $\ln(p/g)$ as a function of the rotational energy gives a linear trend, with the gradient given by $1/(k_{\text{B}} T_{\text{rot}})$, where $k_{\text{B}}$ is the Boltzmann constant.
Repeating this procedure for the rotational level populations averaged over the central $3\times3$ subcells gives the rotational temperature distribution shown in figure \ref{fig:rotationalTemperature}.
The population distribution of each symmetry group is well-described by the Boltzmann distribution; the correlation coefficient of each linear fit is greater than 0.95.

The gas density, and hence the collision rate, is highest in the first sub-cell; most of the rotational cooling occurs in that cell, and over a range too small to be observed in figure \ref{fig:rotationalTemperature}.
The $A$ and $E$ symmetry states do not interconvert during a collision, so the manifold of rotational levels belonging to each symmetry group cool separately at different rates corresponding to the relative cross sections.
In the limit of an infinite number of collisions, the temperature of each state will asymptotically reach the same value, but the relative population of the nuclear spin states will remain at the room-temperature value.
In a molecular beam, however, the density decreases and the velocity increases with distance, so the collision rate decreases, such that after a specific distance no further collisions will occur.
This can be seen in figure \ref{fig:rotationalTemperature}, where the $A$ and $E$ nuclear symmetry states are asymptotically approaching different values, separated by about 4~K.
\textbf{The oscillations visible in the rotational temperature trend at distances beyond about 1~mm are statistical fluctuations arising from the small number of simulated particles within each subcell, and represent the uncertainty in the rotational temperature under these conditions.}

Experimentally, rotational temperatures are estimated by line intensities in rotationally-resolved spectra, measured by pulsed-laser resonance-enhanced multi-photon ionisation (REMPI).
In general, these measurements are made by the laser intersecting the molecular beam several centimetres from the nozzle, after it has passed through a skimmer.
The rotational temperature predicted here, however, is bracketed by rotational temperatures observed experimentally.
Measurements of the skimmed beam 85~mm from the nozzle range from 4~K for a 1\% seed fraction with a 2~bar backing pressure\cite{Tkac2014-State-to-state-resolved-differential} to 40~K for a 5\% seed fraction with 1~bar backing pressure\cite{Ashfold1998-Resonance-enhanced-multiphoton}.

\section{Discussion and Conclusion}

This work demonstrates that (DSMC) is a versatile technique to model collisions in rarefied gases, and extract properties such as temperature and pressure. 
Molecules with rotational degrees of freedom can also change rotational state during a collision, exchanging internal energy with translational energy, and the rotational population distribution eventually comes into equilibrium with the translational temperature of the gas.
We demonstrate that rotational energy transfer can be explicitly included in a DSMC model by a matrix algebra approach, modelling the rotational population transfer as a finite Markov process.
A probability matrix, generated from the rotational state-to-state cross sections, quantifies the redistribution of population from a given rotational level to all permitted final levels.

The supersonic expansion of a beam of NH$_3$ seeded in He from a 0.5~mm orifice into vacuum is a system that exhibits a wide range of density, temperature and collision rates. The modelling of this system confirms the validity of DSMC under free-jet conditions.
Our model accurately reproduces the results of analytical expressions for the velocity and density of a molecular beam over the region of validity of the continuum hydrodynamic approximation used in their derivation.
Closer to the nozzle, we reproduce the high rate of collisions between molecules in the gas expansion.
These collisions accelerate the beam, reducing the translational temperature in the moving frame of reference, and rapidly reduce the NH$_3$ rotational temperature.
Ammonia exists in two nuclear spin symmetry states that do not interconvert during a collision.
These states cool independently at different rates and, owing to a finite number of collisions, reach a different lowest temperature.

In future work, the method employed here will be applied to rotational cooling in buffer gas cells, supporting design improvements to buffer-gas cooled beams for spectroscopy and dynamics experiments.

\section{Funding}

The authors are grateful to the Wiener Anspach Foundation for its financial support of this work. 

\section{Supplemental Data}

Supporting data can be obtained from the Oxford Research Archive, DOI 10.5287/bodleian:ht24wj516.

\bibliographystyle{tMPH}
\bibliography{ref}

\begin{thebibliography}{24}
\providecommand{\url}[1]{\texttt{#1}}
\providecommand{\urlprefix}{URL }
\markboth{Taylor \& Francis and I.T. Consultant}{Molecular Physics}

\bibitem{meijer}
S.Y.T. van~de Meerakker, H.L. Bethlem and G. Meijer,  Nat. Phys.  \textbf{4},
  595 (2008).

\bibitem{merkt}
N. Vanhaecke, U. Meier, M. Andrist, B.H. Meier and F. Merkt,  Phys. Rev. A
  \textbf{75}, 031402 (2007).

\bibitem{osterwalder}
J. Jankunas and A. Osterwalder,  Annu. Rev. Phys. Chem.  \textbf{66} (1), 241
  (2015).

\bibitem{Heazlewood2015-Low-Temperature-Kinetics-and-Dynamics}
B.R. {Heazlewood} and T.P. {Softley},  Annu. Rev. Phys. Chem.  \textbf{66}, 475
  (2015).

\bibitem{Dubernet1989-Rotational-state-dependence}
M.L. {Dubernet} and R. {McCarroll},  Z. Phys. D  \textbf{13}, 255 (1989).

\bibitem{Clary1987-Rate-constants-for-the-reactions}
D.C. Clary,  J. Chem. Soc., Faraday Trans. 2  \textbf{83}, 139 (1987).

\bibitem{Hutzler2012-The-Buffer-Gas-Beam:-An-Intense}
N.R. Hutzler, H.I. Lu and J.M. Doyle,  Chem. Rev  \textbf{112} (9), 4803
  (2012).

\bibitem{Sommer2009-Continuous-guided-beams}
C. {Sommer}, L.D. {van Buuren}, M. {Motsch}, S. {Pohle}, J. {Bayerl}, P.W.H.
  {Pinkse} and G. {Rempe},  Faraday Discussions  \textbf{142}, 203 (2009).

\bibitem{Twyman2014-Production-of-cold-beams}
K.S. Twyman, M.T. Bell, B.R. Heazlewood and T.P. Softley,  J. Chem. Phys.
  \textbf{141} (2), 024308 (2014).

\bibitem{Bird1994-Molecular-Gas-Dynamics-and-the-Direct}
G.A. Bird, \emph{Molecular Gas Dynamics and the Direct Simulation of Gas Flows}
    (Clarendon Press, Oxford, 1994).

\bibitem{Borgnakke1975-Statistical-collision-model}
C. {Borgnakke} and P.S. {Larsen},  J. Comput. Phys.  \textbf{18}, 405 (1975).

\bibitem{Wysong1998-Assessment-of-direct-simulation}
I.J. {Wysong} and D.C. {Wadsworth},  Physics of Fluids  \textbf{10}, 2983
  (1998).

\bibitem{Tokumasu1999-Dynamic-molecular-collision}
T. {Tokumasu} and Y. {Matsumoto},  Phys. Fluids  \textbf{11}, 1907 (1999).

\bibitem{Boyd1993-Relaxation-of-discrete-rotational}
I.D. Boyd,  Phys. Fluids  \textbf{5}, 2278 (1993).

\bibitem{Koura1992-Statistical-inelastic-cross-section}
K. {Koura},  Phys. Fluids  \textbf{4}, 1782 (1992).

\bibitem{Hodges2001-Intermolecular-potential-for-the-interaction}
M.P. Hodges and R.J. Wheatley,  J. Chem. Phys.  \textbf{114}, 8836 (2001).

\bibitem{Gubbels2012-Scattering-resonances-in-slow}
K.B. Gubbels, S.Y.T. van~de Meerakker, G.C. Groenenboom, G. Meijer and A.
  van~der Avoird,  J. Chem. Phys.  \textbf{136} (7), 074301 (2012).

\bibitem{Green1976b}
S. {Green},  J. Chem. Phys.  \textbf{64}, 3463 (1976).

\bibitem{Machin2005-Rotational-excitation-and-de-excitation}
L. Machin and E. Roueff,  J. Phys. B  \textbf{38}, 1519 (2005).

\bibitem{Yang2008-Rotational-excitation-of-NH3-and-ND3-due-to-He-atom}
B.H. Yang and P.C. Stancil,  Eur. Phys. J. D  \textbf{47}, 351 (2008).

\bibitem{Tkac2014-State-to-state-resolved-differential}
O. {Tk{\'a}{\v c}}, A.K. {Saha}, J. {Onvlee}, C.H. {Yang}, G. {Sarma}, C.K.
  {Bishwakarma}, S.Y.T. {van de Meerakker}, A. {van der Avoird}, D.H. {Parker}
  and A.J. {Orr-Ewing},  Phys. Chem. Chem. Phys.  \textbf{16}, 477 (2014).

\bibitem{scholes}
G. Scoles, editor, \emph{Atomic and molecular beam methods}, Vol.~1   (Oxford
  University Press, New York ; Oxford, 1992).

\bibitem{Sanna2005}
G. Sanna, \emph{Introduction To Molecular Beams Gas Dynamics}   (Imperial
  College Press, London, 2005).

\bibitem{Ashfold1998-Resonance-enhanced-multiphoton}
M.N.R. {Ashfold}, S.R. {Langford}, R.A. {Morgan}, A.J. {Orr-Ewing}, C.M.
  {Western}, C.R. {Scheper} and C.A. {de Lange},  Eur. Phys. J. D  \textbf{4},
  189 (1998).

\end{thebibliography}
\end{document}